\documentclass{aa}
\usepackage{times}
\usepackage{graphicx}

\begin{document} 

\title{Growth of supermassive black holes and metallicity in quasars}

\author{Jian-Min Wang}

\authorrunning{J.-M. Wang}

\offprints{Jian-Min Wang \\ \email{wangjm@astrosv1.ihep.ac.cn}}

\institute{Laboratory of High Energy Astrophysics,
Institute of High Energy Physics, The Chinese Academy of Science, Beijing 
100039, and Beijing Astrophysics Center, Beijing University,
Beijing 100871, P.R. China} 

\date{Received 16/July, 2001; accepted 30/July, 2001, by 
                 {\bf Astronomy \& Astrophysics Letter}}

\def\dm{\dot{M}}
\def\nhe{\rm NV/HeII}
\def\Ome{\Omega}
\def\sunm{M_{\odot}}
\def\sunr{R_{\odot}}
\def\sunz{Z_{\odot}}
\def\nc{\rm NV/CIV}

\abstract{The strong correlation between the mass of the central
supermassive black hole (SMBH) and the bulge in some galaxies and 
quasars implies that the formation of the black hole is somehow 
linked to the bulge. The measurement of metallicity by
$\nc$ or $\nhe$ in quasars allows to discuss a possible way of
formation of the black hole. In this {\it Letter} we trace the metallicity
along the possible routes in Rees' diagram in order to test the 
ways by which SMBHs can form. We derive a relation between the 
metallicity and the mass of the SMBH as $Z\propto M_{\rm BH}$ based 
on the numerical simulation of the evolution of star clusters. It is 
in good agreement with the relation determined by the metallicity 
measured by $\nc$ or $\nhe$ in Hamann \& Ferland's sample. This lends 
observational support to the formation of SMBHs via routes R4 or R5, 
namely, the evolution of dense star clusters. %
\keywords{black hole-metallicity-quasars}} 

\maketitle  

\section{Introduction}\label{sect:intro}
There is increasing evidence for the presence of supermassive black holes 
(SMBHs) in the centres of galaxies and quasars from current
observations (see the reviews of \cite{r84}, \cite{r98}, Kormendy \&
Gebhardt 2001). The 
masses of SMBHs in galaxies strongly correlate with the masses of the 
corresponding bulges (\cite{m98}), and even stronger with the
dispersion velocity (\cite{g00}). These facts reflect that
the SMBHs grow with the bulges, albeit in a way which is yet unknown 
(\cite{bs01}). The finding of the relation of Magorrian et al. (1998) 
may give us a clue to understand how SMBHs are formed in galaxies. 
Laor (1998) originally realized that there is a similar relation
between the masses of SMBHs and the bulge luminosities of their host 
galaxies in 14 bright quasars (for active galactic nuclei
see \cite{w99}, \cite{fer01}). In Laor's sample, SMBH masses are 
measured by the velocity dispersion of the H$\beta$-emitting 
clouds, and the bulge luminosities are from {\it Hubble Space
Telescope} ({\it HST}) observations.
These relations are connected to the growth of the SMBHs. However, 
how exactly a black hole grows, remains open.

Rees (1984) gave a map of all possible routes of SMBH formation.
There are mainly two possible ways to form the central SMBH (see the routes 
of SMBH formation in Rees' diagram). The first is that SMBH forms directly 
from the primordial gas cloud, this is further investigated by Loeb
(1993), Loeb\& Rasio (1994). The physical reason is that star
formation would be quenched when the infalling primordial gas reaches 
some critical central concentration (\cite{lr94}).
On the other hand, a dense star cluster may be formed via star formation.
This is supported by the observations of nearby galaxies showing that there 
is a central star cluster with density of  $\sim 10^6 - 10^8\sunm$
and a one dimensional velocity dispersion $\sigma$ of typically 
$\sim 100 - 400$ Km~s$^{-1}$ (\cite{l89}). 
The dense star cluster will inevitably form a SMBH, although the stars
may evolve in somewhat different ways (\cite{br78}, \cite{ds83}, \cite{qs87}).

It has been argued that three parameters, such as the mass and the spin of the
black hole, and the accretion rate, may give a full description of a quasar
(Blandford 1990). Recently, the metallicity, has received attention
as the fourth parameter. It has become evident that quasars have 
very high metallicity, even at very high redshift ($z>4$) (\cite{hf99}).
This means that there must have been very rapid and violent stellar 
evolution in the early stages  of these quasars to produce the 
high metallicity (\cite{hr93}).
Metallicity, as a possible consequence of the formation and growth of SMBHs,
may provide invaluable information of the history of quasars
(\cite{hf99}). 

We are motivated by the chemical evolution of 
quasars based on measurements of metallicity (Hamann \&
Ferland 1992, 1993) to investigate this subject. The observation 
of emission line spectra of quasars can, in principle, provide the 
abundances in the broad line region (BLR), and would allow to deduce 
the details of the chemical evolution of the corresponding nucleus
since the BLR is  the innermost part of a quasar.  In this Letter, we make
an attempt to trace the metallicity of every route in Rees' diagram,
and naturally connect the growth of the SMBH to the observed metallicity
in quasars.  

\section{Observational metallicity in quasars}\label{sect:observation}
High metallicity is common among quasars. As an independent parameter
describing quasars, metallicity may be used to trace the evolutionary 
history of quasars.
The metallicity in quasars may open a new way to understand the growth
of a SMBH. It has been suggested by Hamann \& Ferland (1993) 
that $\nc$ or $\nhe$ are very good indicators of the metallicity in quasars.
They subsequently measured the metallicity in a large sample of quasars 
(Hamann \& Ferland 1993, 1999). There appears unambiguous evidence for 
a correlation between metallicity and luminosity. Although we do not
fit this correlation via regression, it is found that
\begin{equation}
\left({\rm \frac{NV}{CIV}}\right)\propto \left(\nu L_{\nu}\right)^{\alpha_1},
{~~\rm or~~\left(\frac{NV}{HeII}\right)}\propto 
              \left(\nu L_{\nu}\right)^{\alpha_2},
\end{equation}
where $\alpha_1 \approx \alpha_2\approx 0.5$, and $L_{\nu}$ is the specific 
luminosity at frequency $\nu$. The reverberation technique provides a set 
of available data of quasars (Kaspi et al 2000), and shows a relationship
between the mass of a black hole and its luminosity
$M_{\rm BH}\propto \left(\nu L_{\nu}\right)^{\beta}$ with
$\beta\approx 0.5$. From the above relations, we have
\begin{equation}
{\rm \left(\frac{NV}{CIV}\right)}\propto M_{\rm BH}^{\alpha_1/\beta},
{~~\rm or~~\left(\frac{NV}{HeII}\right)}\propto M_{\rm BH}^{\alpha_2/\beta}.
\end{equation}
This relation is mainly derived observationally, and the only assumption is 
that the BLR clouds are virialized in the potential of the SMBH. It is 
generally believed that this approximation works in quasars (Netzer 1990).

\section{Metallicity in Rees' diagram}\label{sect:rees}
Rees (1984) lists the possible ways to form SMBHs in the universe. There
are five possible routes. Route 1 (R1) is collapse and/or accretion directly
from an original gas cloud. Route 2 (R2) is through post-Newtonian
instability of a supermassive star. The supermassive star may be formed
through two possible ways: 1) direct collapse from a gas cloud,
2) from collisional disruption of stars in a dense star formed
from the gas cloud. Route 3 (R3) is the merger of a black hole binary 
formed from the supermassive star induced by bar-mode instability. 
Route 4 (R4) means that a SMBH is born through spectacular accretion
in a cluster of neutron stars or stellar-mass black holes. Route 5 (R5)
suggests that a SMBH is formed in a relativistic cluster of compact objects
as in R4, but the relativistic instability or gravitational radiation will
induce the formation of SMBH.

First we check route 1 (R1). This route was studied by Loeb \& Rasio 
(1994). In such a scenario, the SMBH is formed directly due to a catastrophic
collapse or it grows via gradual accretion. As we show below, a supermassive
black hole may be formed via a single catastrophic collapse, but without 
ejection of metals since the gravitational binding energy is too strong to 
eject matter. Thus there will be no relation between metallicity and
the mass of the black hole.

A supermassive star is first formed via two possible ways before collapsing
to a SMBH. According to R2 a SMBH will be formed due to a post-Newtonian 
instability, R3 suggests that the SMBH 
is formed due to the merger of binary black holes. This process will 
lead to huge radiation of gravitational waves. It may be confirmed by the 
detection of gravitational radiation. The two routes (R2 and R3) employ 
the formation of a supermassive star. Here we will show that there is an 
upper limit for the mass of a SMBH created via R2.

The structure and evolution of the supermassive star is insufficiently 
understood. We assume that the supermassive star is composed of hydrogen.
The gravitational binding energy of an object with mass $M$ and 
radius $R$ is $E_{\rm g}\approx GM^2/R$, 
where $G$ is the gravitational constant.  
Its total nuclear energy is $E_{\rm n}=\epsilon Mc^2$ with 
$\epsilon$  being the conversion efficiency ($\epsilon=0.007$ for the 
burning of hydrogen to helium) and $c$ is the speed of light. If we adopt 
the mass-radius relation $R=R_0 \left(M/\sunm\right)^{q}$ for 
supermassive stars, where $R_0=9.0\times 10^{11}$~cm and
$q=0.47$ (Collins 1989), we will have the 
upper limit of the mass of the supermassive object ejecting 
matter during the supernova explosion
\begin{equation}
\frac{M_{\rm c}}{\sunm}= 
(2\epsilon)^{1/(1-q)}\left(\frac{c}{v_0}\right)^{2/(1-q)}
=4.5\times 10^{6},
\end{equation}
based on $E_{\rm g}\leq E_{\rm n}$,
where $v_0=\left(G\sunm/R_0\right)^{1/2}$. No mass ejection takes 
place during the formation of the black hole, albeit the collapse of the 
supermassive star for $M\geq M_{\rm c}$.
If a SMBH forms directly from the collapse of a primordial gas cloud, 
as the mass of SMBH is much higher than $M_{\rm c}$,
the detectable metallicity may be low 
because (1) no heavy elements produced in the collapse have been
ejected, and (2) the dominance of radiation pressure in supermassive stars 
and compact supermassive disks would prevent fragmentation and star 
formation in these systems (Loeb 1993, Loeb \& Rasio 1994). 
Although a strong wind may developed from the suface of supermassive 
star, the first reason (1) is still supported by the fact that the 
timescale of contraction of the supermassive star is much shorter than 
that of the mass loss through the wind on the surface of supermassive star.
The contraction timescale of a supermassive star approximates to 
$t_{\rm contr}\approx 3\times 10^3R_{14}(M/M_{\rm c})^{1/2}$~yr, where
the typical dimension of the supermassive star $R_{14}=R/10^{14}$~cm
(\cite{f73}). This timescale is much shorter than the mass loss
timescale $t_{\rm w}=M_{\rm c}/\dot{M}\approx 10^{10}(M/M_{\rm c})$~yrs 
(\cite{bond84}). Therefore there is no significant ejection of metals 
during the rapid evolution of the supermassive star.

It has been argued that the collision between a normal galaxy and a naked 
SMBH leads to the activity in a galaxy (\cite{ft96}), but in this case
there is no correlation between metallicity and the mass of SMBH. On the 
other hand, if a SMBH is formed in a dense star cluster via coalescence 
of stars or by a merger of stellar mass black holes, there is an approximate 
proportional relationship between the metallicity and the mass of the SMBH 
according to the dynamical and stellar evolution of the cluster. 
We will derive a crude approximation about this relation below.

Routes 4 and 5 (R4 and R5) suggest that the SMBH is formed in a dense star
cluster. In R4 and R5, a cluster with stars heavier than 100$\sunm$
is formed in advance. The high rate of supernova explosions leads to the
formation of a cluster of neutron stars or steller-mass black holes. We 
assume that some of the ejected medium from the supernova explosions may 
form clouds in the broad line region. The two ways have a rough 
metallicity prediction as we show in the
following. Following Duncan \& Shapiro (1983), we assume that
the cluster consists of $N=10^8N_8$ stars with a dispersion
velocity $v=350v_{350}({\rm Kms^{-1}})$, this star cluster thus has
a core with a radius $R_{\rm c}=2.3N_8v_{350}^{-2}$pc.
Out of the radius $R_{\rm c}$, the initial density has a power law profile 
$n(r)\propto r^{-p}$, here $p$ is the index. This power law distribution 
may work further into the klio parsec scale bulge (\cite{b01}).
The two-body relaxation time is 
$t_{\rm r}=1.5\times 10^{10}N_8\Lambda_{18}^{-1}v_{350}^{-3}$yr, where
$\Lambda_{18}=\ln (0.5N)/18$ and the initial masses
of all stars are taken as one solar mass $m$. The total mass and the collision
time scale of star cluster are $mN$ and 
\begin{equation}
t_{\rm coll}=6.8\times 10^{10}N_8^2v_{350}^{-5}(1+1.3v_{350}^2)^{-1}~
{\rm yr},
\end{equation}
respectively.
The black hole grows due to the coalescence of stars via collision and
tidal capture of stars because the collisional interactions are inelastic. 
Detailed numerical simulations show that the growth of a black hole 
can be divided into two phases for an isothermal sphere of solar 
type stars in the absence of a black hole (Duncan \& Shapiro 1983):
\begin{equation}
\dot{M}_{\rm BH}(M_{\odot} \rm yr^{-1})=\left\{
                 \begin{array}{ll}
                  g_0, &{\rm for~~} 0 \le t \le t_{\rm coll}\\
                  g_0t^{-\gamma},& {\rm for~~}t\geq t_{\rm coll}.
                  \end{array}\right.
\end{equation}
where $g_0$ is  a constant related to $N_8$ and $v_{350}$,  
$\gamma=(3-2p)/p\approx (0.9\sim 1.0)$ is the growth index in the later 
phase weakly depending on the cluster parameters. For example,
$g_0\approx 1.7$, $\gamma=1$ and $t_{\rm coll}=2.0\times 10^8$yr for
the case of $N_8=2.7$, $v_{350}=2.9$ (Duncan \& Shapiro 1983).  
Here we neglect the small peak 
of $\dot{M}_{\rm BH}$ around the critical time $t_{\rm coll}$. After 
the collisional phase the growth of the black hole is mainly via the capture 
of stars from the cluster. It is thus expected that the main phase of 
metallicity enrichment is within the the time $t_{\rm coll}$.

The Kelvin-Helmholtz time scale approximates to
$t_{\rm KH}\approx 3.0\times 10^7(m/M_{\odot})^{-2}$yr, which represents 
the time scale to form a new star after coalescence (Quinlan \& Shapiro 
1987). Numerical 
results show that a massive star with about 50$M_{\odot}$ can grow 
within a time scale $t_{\rm coal}\approx 10^6$~yr due to colliding star
coalescence if the dispersion velocity is less than its escape
velocity (Colgate 1967). The maximum mass of a star due to coalescence via 
collision can be estimated from the fact that the coalescence process 
saturates when an ordinary $1\sunm$ star cannot be captured by the 
coalescencing star since this star can pass straight through them. 
Supposing the
mass-radius relation to be $r^*/\sunr=\left(m^*/\sunm\right)^{3/4}$,
 we have
\begin{equation}
m_{\rm max}^*\leq \left(\frac{v}{v_{\rm esc}}\right)^8\sunm
              =53\left(\frac{v_{350}}{2.9}\right)^8\sunm,
\end{equation}
where $v_{\rm esc}=(2G\sunm/\sunr)^{1/2}$.
This mass limit is sensitive to the dispersion velocity. Considering 
some other physical rules, the coalescencing stars may grow up to 
50 $\sim$ 100 solar mass (\cite{br78}). 

The time scale for star evolution approximates to 
$t_{\rm evo}=6.0\times 10^6(50M_{\odot}/M)$yr for stars larger than 
$12M_{\odot}$. $t_{\rm KH}$ is much shorter than the coalescence and 
evolution time scales of massive stars. It appears that the solar mass 
stars in the cluster may form many massive stars following the collisions 
and coalescence within the collision time scale. Once a massive star forms, 
the ejection of metal-rich ejecta will follow the post-main 
sequence nuclear burning, culminating most probably in a Type II supernova. 
The heavy $(Z>6\sunz)$ element mass $M_{_{\rm Z}}$ yield in ejecta  varies with the 
progenitor mass, but is expected to be in the range of $4\sim 40$ 
$M_{\odot}$ (\cite{ww86}). The rapid rotation of the progenitor 
can modify the metal yield by a modest amount as shown by the
calculations (\cite{bw83}). Although it remains uncertain, the most likely 
remnants of massive progenitors are stellar-mass black holes with 
$\sim $ 10 solar mass (\cite{ww86}). Considering that the life time of
a massive star is much shorter than the collisional time scale, the total 
mass of metals produced by supernova explosion is about
$M_{\rm Z}=m_{_{\rm Z}}\left(dN/dt\right)t$, then the metal abundance $Z$
can be obtained by,
\begin{equation}
Z\approx \left(\frac{m_{_{\rm Z}}}{m}\right)
 \left(\frac{1}{N}\frac{dN}{dt}\right)t
  =\left(\frac{m_{_{\rm Z}}}{m}\right)\left(\frac{t}{t_{\rm coll}}\right).
\end{equation}
Combining eqs (5) and (7) we obtain the following relation between
the mass of a growing black hole and the metallicity as
\begin{equation}
Z=\frac{1}{g_0t_{\rm coll}} \left(\frac{m_{_{\rm Z}}}{m}\right)M_{\rm BH},
~~~{\rm for~~} 0\leq t \leq t_{\rm coll}.
\end{equation}
When $t\geq t_{\rm coll}$, as the growth of the SMBH is dominated by 
the capture of stars, the above relationship will not hold. 
Detailed numerical calculations using a photoionization model
show that the abundance $Z$ is roughly proportional to the two ratios,
\begin{equation}
Z\propto {\rm \left(\frac{NV}{CIV}\right)^{\gamma_1}},
{~~~\rm or~~}Z\propto {\rm \left(\frac{NV}{HeII}\right)}^{\gamma_2},
\end{equation}
and $\gamma_1\approx \gamma_2\approx 1$ measured from
the right panel in Figure 6 of Hamann \& Ferland (1999). We would like
to point out that this result is based on a series of broad emission line 
simulations that hold all other parameters fixed while the abundances are
varied. We should keep this in mind. 
Thus the observable metallicity is given by
\begin{equation}
{\rm \left(\frac{NV}{CIV}\right)}\propto M_{\rm BH}^{1/\gamma_1},
{~~~\rm or~~\left(\frac{NV}{HeII}\right)}\propto 
M_{\rm BH}^{1/\gamma_2}.
\end{equation}
This formula works within the collisional timescale $t_{\rm coll}$ 
which is determined
by the cluster itself. We find that the theoretical
prediction (10) is in good agreement with the observational relation (2).
Therefore routes 4 and 5 (R4 and R5) may be the likely ways to form
a SMBH from a primordial gas cloud.

\section{Conclusions and discussions}\label{sec:conclusions}
In this paper we trace the metallicity along the routes in Rees'
diagram in order to test the possible ways by which SMBHs form in quasars. 
The argument that a single collapse of a supermassive star forms a SMBH 
is not supported by the observed relation between metallicity and 
luminosity in quasars. The reason is that the mass is too high for the 
ejection of metals since its self-gravitation energy is larger than the 
energy of nuclear reaction. We derive a formula of growth of a black 
hole with metallicity, which is in good agreement with
the observed relation between metallicity and luminosity. Thus 
observations seem to support the formation of SMBHs
from a dense star cluster via coalescence and evolution of stars.

There is a modified route 1 (hereafter MR1) suggested by the 
referee. It states that if stars form during the collapse, they
will enrich the gas with metals before the formation of the central SMBH.
The gas that forms the SMBH might be processed by a generation of stars in
the host galaxy before it gets funneled into the core to make the SMBH.
Such a complex formation of SMBH depends on many processes, fragmentation of
cloud during the collapse, the formation of multi-generations of stars.
Let us speculate that a primordial gas cloud with mass of several 
$10^8\sunm$ is gravitationally fragmentated into several hundreds of
massive stars with mass less than 4.5$\times 10^6\sunm$. These massive
stars evolve undergoing hydrogen burning, then eject metals due to 
supernovae explosion, leaving a cluster with several hundreds of 
intermediate mass black holes ($<10^6\sunm$). What's the observational 
consequence of such a cluster? This interesting route will be explored 
in the future.

Recently, much attention has been given to the recurrent activity model 
of quasars (\cite{hr93}), which seems to be supported by observations
(\cite{k01}). The mass growth of a SMBH in the first generation 
of quasars may be mainly due to R4 or R5. The masses of SMBHs in 
recurrent quasars may not increase significantly in the
subsequent generations within the Eddington timescale 
$M_{\rm BH}/\dot{M}_{\rm Edd}\approx 5.0\times 10^8$ years.
This argument might be supported by the fact that
the relation of the SMBH mass with the bulge mass in quasars (\cite{l98})
is similar to the Magorrian et al. relation in galaxies.
We note that the slope of the relation found by Laor (1998) is slightly
different from that of the Magorrian et al. relation.
If the differences are real, we might then deduce the
evolution of SMBH accretion after the first generation. This is 
worthy of a future investigation.

\begin{acknowledgements}
The author is very grateful to an anonymous referee for the stimulating
suggestions to improve the manuscript. Professor R. Staubert is highly
acknowledged for a careful reading of the manuscript to improve
the English and helpful suggestions. He appreciates the helpful 
discussions with Professors A. Laor, A. Wandel, H. Netzer,
Z. Han and Dr. O. Shemmer. The author is supported by the 
``Hundred Talents Program'' of The Chinese Academy of Sciences. This 
research is financed by the Special Funds for Major State Basic 
Research Project.
\end{acknowledgements}

\end{document}